\documentclass[final,authoryear,3p,times,twocolumn]{elsarticle}
\pdfoutput=1
\usepackage{ecrc}
\usepackage{natbib}
\usepackage{graphics}
\usepackage{amssymb}
\usepackage{amsmath}
\usepackage{lineno}
\biboptions{square}

\volume{00}
\firstpage{1}
\journalname{Nuclear Instruments and Methods in Physics Research A}
\runauth{}
\jid{NIM A}
\jnltitlelogo{Nuclear Instruments and Methods in Physics Research A}
\begin{document}

\begin{frontmatter}

\title{Scintillation decay time and pulse shape 
discrimination in oxygenated and deoxygenated solutions of linear 
alkylbenzene for the SNO+ experiment}
\author{H.\,M.\,O'Keeffe\corref{cor1}}
\author{E.\,O'Sullivan}
\author{M.\,C.\,Chen}
\cortext[cor1]{Corresponding author: okeeffe@owl.phy.queensu.ca}
\address{Department of Physics, Engineering Physics and Astronomy, 
Queen's University, Kingston, Ontario, K7L 3N6, Canada}
\begin{abstract}
The SNO+ liquid scintillator experiment is under construction in the
SNOLAB facility in Canada.  The success of this experiment relies upon 
accurate characterization of the liquid scintillator, linear alkylbenzene 
(LAB).  In this paper, scintillation decay times for alpha and electron 
excitations in LAB with 2 g/L PPO are presented for both oxygenated and 
deoxygenated solutions.  While deoxygenation is expected to improve
pulse shape discrimination in liquid scintillators, it is not commonly
demonstrated in the literature.  This paper shows that for linear 
alkylbenzene, deoxygenation improves discrimination between electron and 
alpha excitations in the scintillator.
\end{abstract}
\end{frontmatter}
\section{Introduction} 
\label{intro} 
The SNO+ experiment is currently under construction in the SNOLAB facility, 
located approximately 2 km underground in Sudbury, Ontario, Canada. The 
detector will consist of $\sim$780 tonnes of linear alkylbenzene (LAB) liquid 
scintillator held in a 12 m diameter acrylic sphere and surrounded by 7000 
tonnes of ultra pure light water shielding. An array of $\sim$9500 
photomultiplier tubes will detect scintillation light produced by particle 
interactions. Both the acrylic vessel and PMT array were inherited from the 
Sudbury Neutrino Observatory (SNO) experiment \cite{SNO}. 

The SNO+ physics programme will include measurements of low energy solar 
neutrino fluxes and a search for neutrinoless double beta decay using 
neodymium \cite{SNO+}.  Sensitivity to both requires precise energy and 
position reconstruction, for which an accurate characterization of the 
scintillator timing is needed. The scintillator decay time can be used to 
discriminate between alpha and electron events which occur in the 
scintillator, helping to exclude backgrounds and further improve sensitivity. 

The focus of this paper is the measurement of the timing profile of liquid 
scintillator that will be used in the SNO+ experiment, namely linear 
alkylbenzene with 2 g/L 2,5-diphenyloxazole (PPO). Linear alkylbenzene was 
chosen because of its high flash point, low toxicity and acrylic 
compatibility.  PPO will be added to the LAB at a concentration of 2 g/L as 
the primary fluor, emitting scintillation light across a wavelength region 
in which the SNO+ photomultiplier tubes are most efficient. 

Measurements of scintillator decay times for electron and alpha excitations 
were made for oxygenated and deoxygenated samples of LAB with 2 g/L PPO.  
These measurements were made using the single photon sampling technique 
\cite{Bolli}.  This method has been used to derive the 
decay times of many common liquid scintillators, including pseudocumene 
\cite{BOREX}. This paper will briefly discuss the theory of the single 
photon counting method and the experimental apparatus used in this work. 
Finally, results for each configuration of LAB will be presented and 
conclusions drawn. 

\section{Scintillator timing profile} 
When an ionizing particle enters the scintillator, it can excite a linear 
alkylbenzene molecule which non-radiatively transfers this energy to a PPO 
molecule.  Scintillation light is then emitted over a finite time period, 
via radiative de-excitation of the excited PPO molecule.  The timing profile 
is a measure of the intensity of scintillation light as a function of time 
due to a single event.  Timing profiles for organic 
scintillators usually consist of several exponential decay components. 
Radiative transitions to the ground state are permitted from singlet, but not 
triplet, excited states.  The light produced by the singlet state 
de-excitation occurs quickly and is associated with the fastest component in 
the timing profile.  When excitation to a triplet state occurs, a number of 
light producing de-excitation channels are available.  For example, two 
triplet state molecules could collide, 
allowing simultaneous population of the singlet excited state and decay to the 
ground state.  When compared with the singlet state de-excitation, such 
processes occur over a longer time period and are therefore associated with 
the longer timing profile components.  The shape of the timing profile is 
related to the ionization density of the charged particle interacting with the 
scintillator.   The relative contribution of singlet and triplet states to the
timing profile depends on the ionization density of the particle.  This 
results in differences in the shape of the scintillation light waveform which 
can be used to discriminate between particle types.  Of particular importance 
for the SNO+ experiment is discrimination between alpha and electron events 
in the scintillator.  Both solar neutrino and neutrinoless double beta decay
events will produce electron-like signals in the detector.  The ability to
identify alpha events enables their rejection. Since alpha particles produce 
high ionization density, the proportion of the fast component relative to the 
slow component is reduced because of ionization quenching \cite{Leo}.  
Conversely, due to their lower ionization denisty, the timing profile of 
electron events is dominated by the fast component.

\section{Experimental measurement of the timing profile}
\label{sample} 
The timing profile can be measured using the single photon sampling discussed 
in \cite{Bolli}.  In this method, the light produced by a sample of 
scintillator is observed by two photomultiplier tubes (PMTs) connected to fast 
timing discriminators.  The first PMT observes all events and provides a 
trigger for the electronics chain.  The second PMT is covered by a mask 
containing a small hole in the centre, which allows single photons to be 
detected.  The unmasked PMT provides a reference (start) time for each 
event.  A finite time later, a single photon is detected by the masked 
PMT which defines the end of an event.  The time difference between the 
start and stop of each event is recorded.  By producing a histogram of these 
delayed coincidence events, accounting for the background and timing 
resolution, the timing profile of the scintillator can be obtained.

\begin{figure*}[t!] 
\centering 
\includegraphics[totalheight=0.30\textheight]{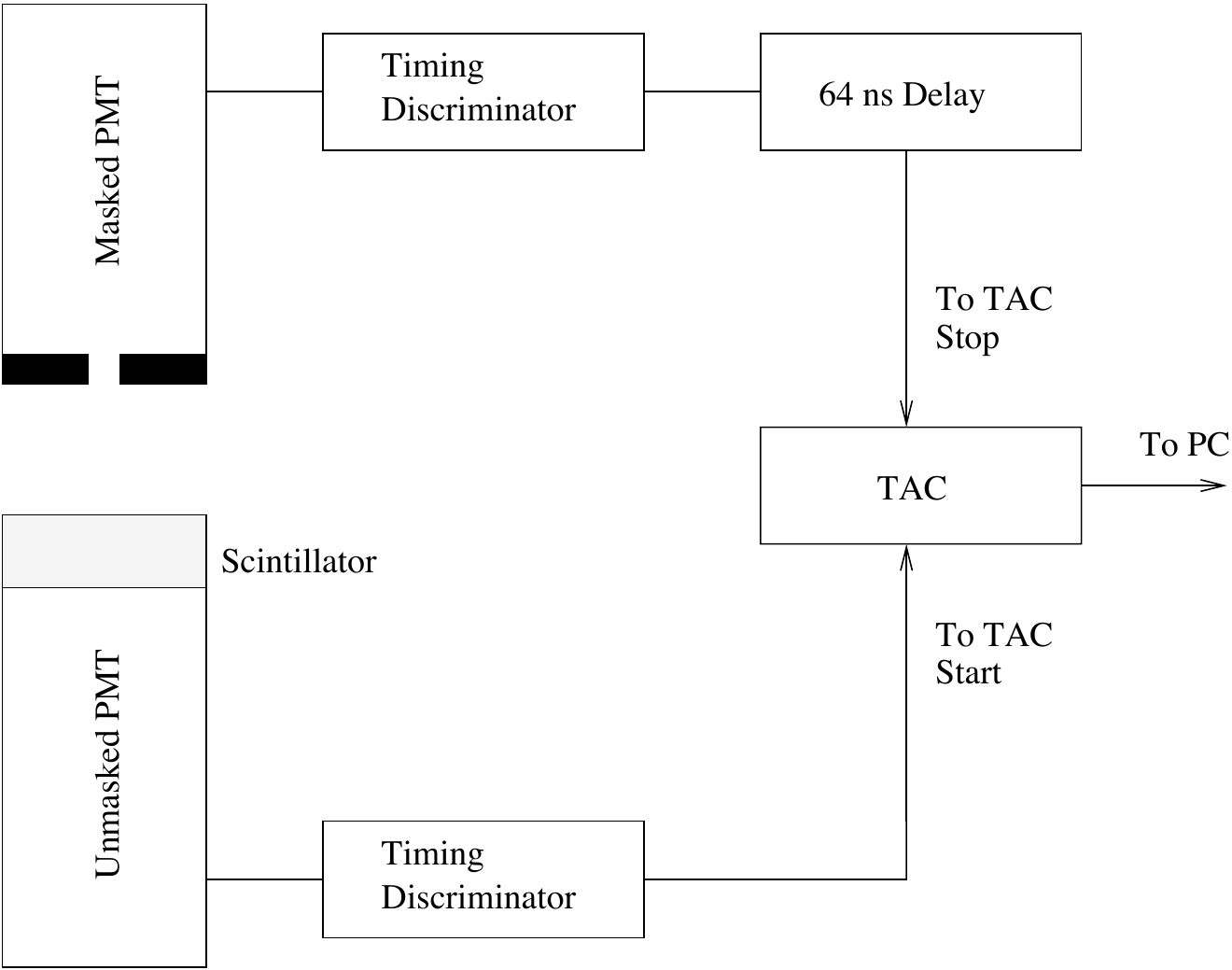} 
\caption{Schematic of the electronics set up for timing profile measurements. 
\label{elec}} 
\end{figure*}

A schematic diagram of the apparatus used in this work is shown in Figure 
\ref{elec}.  A glass dish containing $\sim$50 ml of LAB with 2 g/L PPO 
scintillator was optically coupled to a 5 cm diameter PMT (Electron Tubes Ltd 
9266KB). The second PMT (Electron Tubes Ltd 9266KB) was covered with a mask 
containing a $\sim$1 mm diameter hole.  The signal from each PMT was connected 
to an independent channel of a fast timing discriminator (Phillips 715).  The 
discriminator threshold for each PMT was set by using the discriminator signal 
to gate an MCA energy spectrum.  For the masked PMT, the discriminator 
threshold was set such that the energy spectrum cut off just below the single 
photoelectron peak for all runs.  For the unmasked PMT, the discriminator
threshold was set such that a suitable energy threshold was selected, which was
the same for alpha and electron runs.  This ensured the zero offset of the
timing spectrum was identical for all runs.  The output from the unmasked PMT 
discriminator was connected to the start input of a time to amplitude 
converter (TAC) (ORTEC 566). The masked PMT discriminator channel was 
connected to the stop input of the TAC. A range of 500 ns was used for the 
TAC time window. The output of the TAC was connected to a PC running the 
Maestro multichannel analyzer (MCA) data acquisition software, which 
recorded the time difference for each event in terms of MCA channels. 

Cesium-137 was used to obtain a sample of electrons via Compton scattering of 
the 662 keV gamma.  The discriminator threshold was set to accept the full
Compton edge.  An americium-241 source was used to produce a sample of 
alpha events.  This source was immersed in the scintillator and the 
discriminator threshold was set to exclude the 59 keV gamma, but include the 
full 5.48 and 5.44 MeV (quenched) alpha peaks.  

Coincidences between the PMTs could be caused by non-scintillation events.  To 
characterize this, complementary background runs were taken before and after 
each timing run. In a background run, the radioactive source remained in place 
and the hole in the mask was covered to prevent detection of photons from 
scintillation.  The background spectrum was dominated by fast coincidences 
from cosmic ray interactions in the PMT glass and random coincidences 
between dark noise events.  Each background run was time normalized and 
subtracted from its corresponding timing run.  

\section{Timing calibration and resolution} 
\label{time} 
The timing resolution was measured by removing the PMT mask, which allowed 
scintillation light to be recorded by both PMTs simultaneously.  This produced 
a Gaussian distribution of timings, with sigma equal to the timing resolution 
of the apparatus.  Delays ranging from $\sim$5 ns to $\sim$100 ns were 
introduced between the discriminator and the stop channel of the TAC.  This 
additional delay shifted the mean of the Gaussian by a given number of MCA 
channels and the timing calibration was obtained by applying a linear fit to 
this data.  A timing resolution of $1.9 \pm 0.2$ ns was obtained by fitting a Gaussian 
distribution to this data along with a conversion factor of 
$16.98 \pm 0.85$ MCA bins per nanosecond. 

\section{Results} 
\label{res} 
\begin{figure*}[t!] 
\centering 
\includegraphics[totalheight=0.35\textheight]{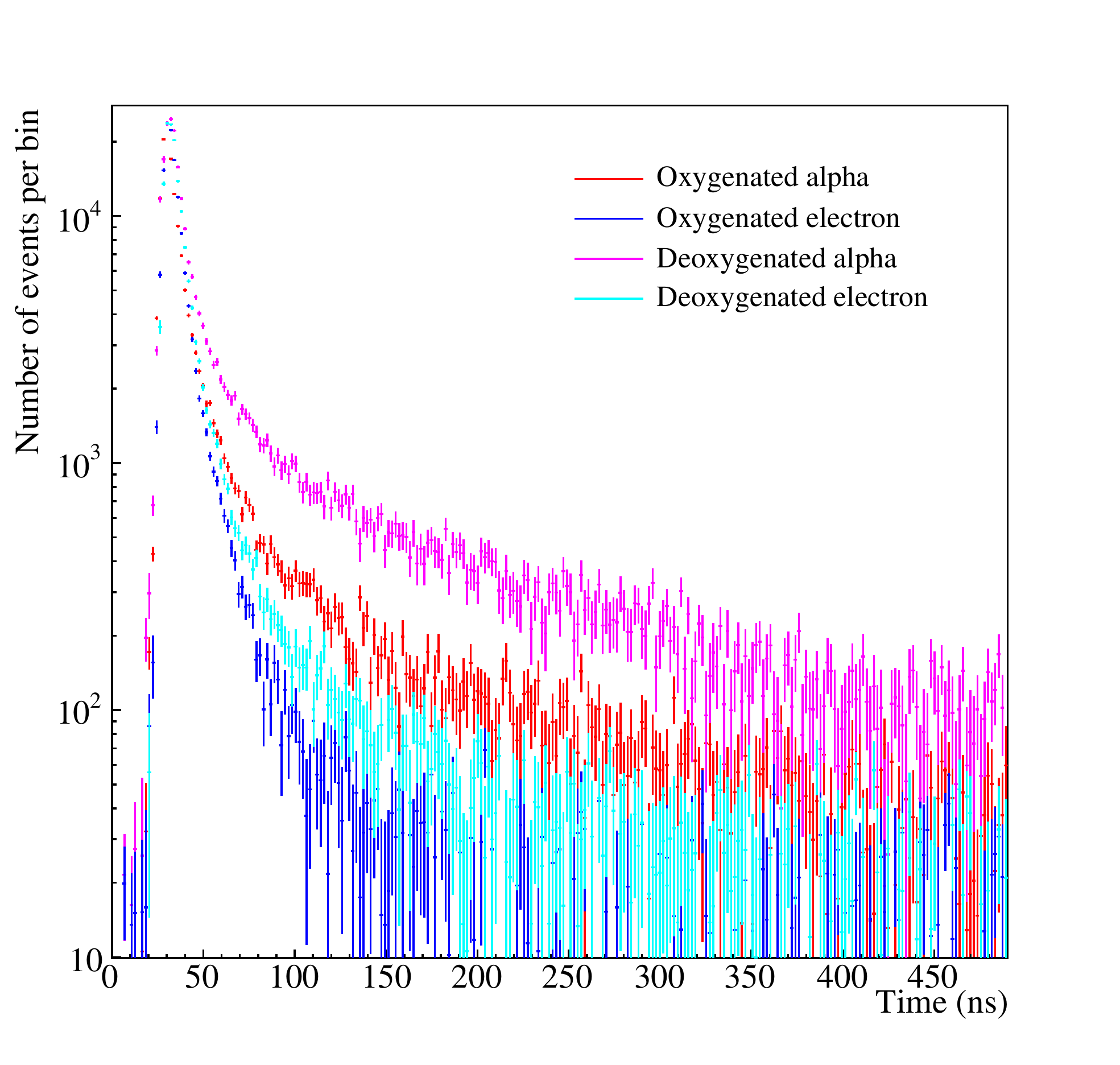} 
\caption{Alpha and electron timing profiles for oxygenated and deoxygenated 
samples of LAB + 2 g/L PPO. \label{aboxdeox}} 
\end{figure*}

Measurements of scintillator decay times for electron and alpha excitations 
were made for oxygenated and deoxygenated samples of LAB+2g/L PPO.  
Deoxygenated samples were prepared by bubbling dry nitrogen through the 
scintillator for 20 minutes.  To ensure the sample remained free of oxygen for 
the duration of the experimental run, the PMTs, scintillator and source were 
enclosed in an acrylic housing through which a slow flow of nitrogen was 
maintained.  Timing profiles for the deoxygenated and oxygenated mixtures are 
shown in Figure \ref{aboxdeox}.  

The timing profiles shown in Figure \ref{aboxdeox} have been peak normalized 
to facilitate comparison between the two modes of excitation.  The first 
portion of the curve is similar for all curves, but those for alpha 
excitations appear to fall off more slowly indicating, as expected, the 
presence of a long tail component. 

To extract the timing components, a function consisting of three decaying
exponentials which were individually convolved with a Gaussian was fitted to the
background subtracted data. Fits with fewer exponentials returned a poorer chi 
squared per degree of freedom and those with more resulted in duplication of 
timing components.  Therefore, three exponential terms were used in each fit. 
Each convolved exponential was multiplied by a scaling factor which represented 
the weighting of that component in the fit.  The function fitted to the data 
was

\begin{equation} 
\label{fit} 
\sum_{i=1}^{3}  A_i \exp\Bigg({\frac{x}{t_i}} + {\frac{0.25}{\sigma {t_i}^2}} \Bigg)
\sqrt{\frac{\pi}{4\sigma}}\Bigg[1+Erf \Bigg(\sqrt{\sigma}
\Bigg(-x-\Bigg(\frac{0.5}{t_i\sigma} \Bigg)\Bigg)\Bigg)\Bigg]
\end{equation} 
where $A_i$ is the scaling factor for each exponential component, $t_i$ is the 
timing component in MCA bins, $x$ is the MCA bin and $\sigma$ is the timing 
resolution (equivalent to 1.9 ns). The relative contributions of each component 
is given by 

\begin{equation} 
\label{rel} R_i = \frac {A_i t_i}{\sum_{i=1}^{3} A_i t_i} 
\end{equation} 
where $R_i$ is the relative contribution for the $i$th component of the fit and 
$A_i$, $t_i$ were derived from \eqref{fit}.  The TF1 function fitting class, 
found in the ROOT analysis framework was used to perform the fit to the data.  
A cross check was made using the ROOFIT software also found in ROOT.  Results 
from the TF1 fit are shown in Table \ref{timing_sum}. 
\begin{table*}[ht!] 
\begin{center} 
\begin{tabular}[ht!]{lcccc} 
\hline 
 & Oxygenated $\alpha$ & Oxygenated $e^-$ & Deoxygenated $\alpha$ & Deoxygenated $e^-$ \\
\hline
$t_1$ (ns) & $4.4 \pm 0.2$   & $4.3 \pm 0.3 $  & $3.2 \pm 0.2$   & $4.6 \pm 0.3$\\
$t_2$ (ns) & $20 \pm 1 $     & $16 \pm 1$      & $18 \pm 1$      & $18 \pm 1$ \\
$t_3$ (ns) & $178 \pm 10 $   & $166 \pm  11$   & $190  \pm 10$   & $156 \pm 9$ \\
\hline
$A_1$      & $520 \pm 6$     & $768 \pm 12$    & $794 \pm 7 $    & $753 \pm 14$ \\
$A_2$      & $59  \pm 3$     & $59  \pm 4$     & $53 \pm 3$      & $61 \pm 3$ \\
$A_3$      & $3.3 \pm 0.1$   & $0.8 \pm 0.1$   & $12.6 \pm 0.2$  & $2.2 \pm 0.1$ \\
\hline
$R_1$ (\%) & 55              & 75              & 44              & 71\\
$R_2$ (\%) & 28              & 22              & 16              & 22\\
$R_3$ (\%) & 17              & 3               & 41              & 7\\
\hline
\end{tabular} 
\caption{Summary of timing results for the alpha and electron timing profile 
fits.  Total errors are given. \label{timing_sum}} 
\end{center} 
\end{table*} 

Results for the timing components show significant differences between alpha 
and electron timing profiles for both the oxygenated and deoxygenated 
scintillators.  Oxygen is responsible for quenching the longer lived decay 
processes. Deoxygenating the scintillator reduces the quenching and thus 
increases the relative amount of the longest timing component.  It is clear 
that deoxygenation has a greater effect on the timing profile of alpha events. 
This enhances differences between the timing profiles of electron and alpha 
scintillation events, implying that pulse shape discrimination should be more 
effective in deoxygenated samples of linear alkylbenzene. 

\begin{figure*}[t!] 
\centering 
\includegraphics[totalheight=0.35\textheight]{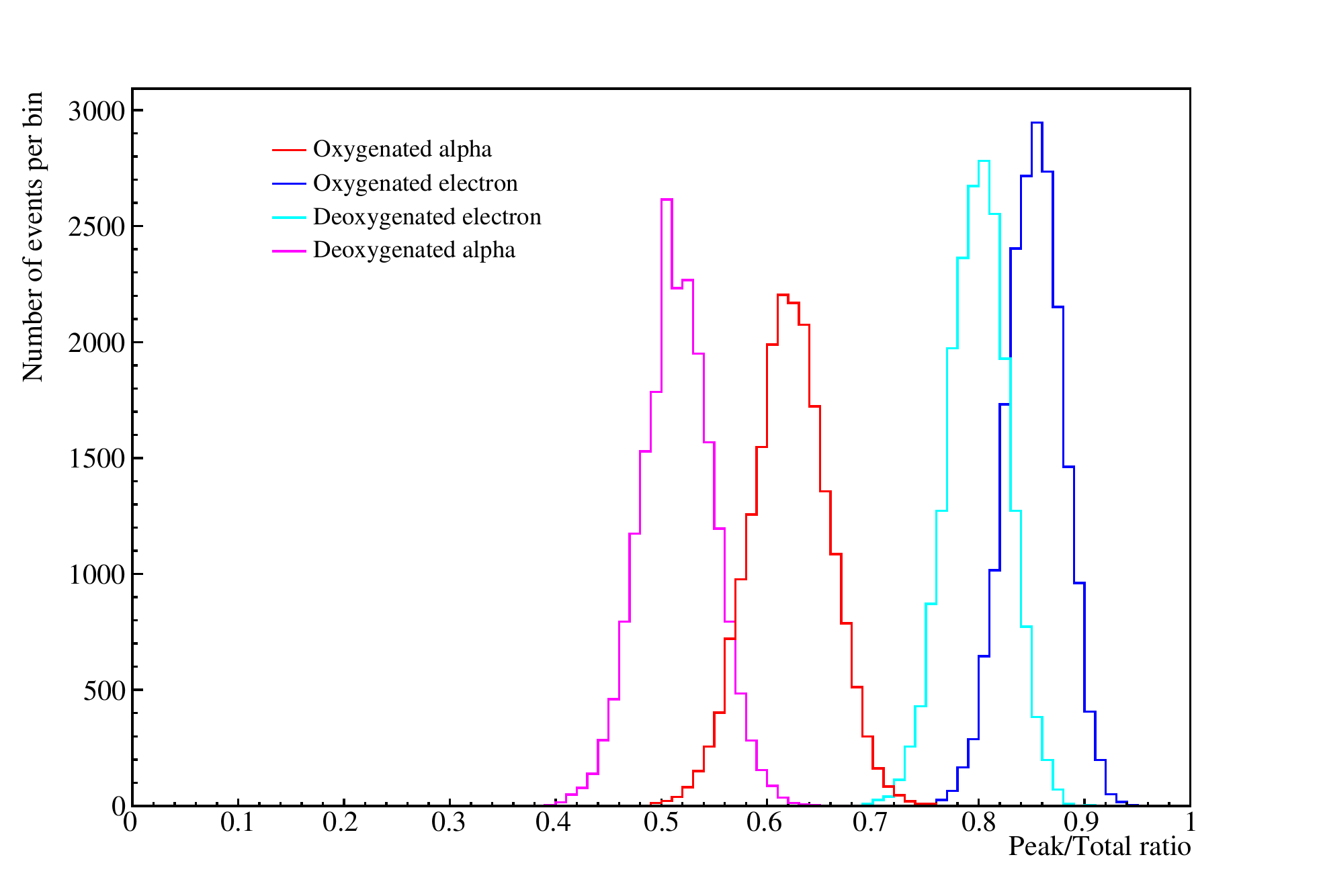} 
\caption{Peak to total distributions for excitations due to electrons and 
alpha particles in oxygenated and deoxygenated LAB + 2 g/L PPO.
\label{tailtototal}} 
\end{figure*}

By defining a peak region in the timing distribution and comparing the 
integral of this region to the total integral, a peak to total ratio could be 
calculated. To assess the difference between the calculated ratios for alpha 
and electron events, 300 points were randomly sampled from each timing 
distribution, which is approximately equal to the number of photons produced 
by a 5 MeV alpha or 0.5 MeV electron. The ratio was calculated by dividing 
the number of counts in the peak region by the total counts. Looping over 
this method many times gave a Gaussian spread. Ratios for both 
oxygenated and deoxygenated scintillators are shown in Figure 
\ref{tailtototal}. 

In oxygenated linear alkylbenzene, approximately 99\% of alphas can be 
rejected whilst retaining $>$99\% of electrons.  In the case of 
deoxygenated linear alkylbenzene and for the same peak region, $>$99.9\% 
of alphas can be rejected whilst retaining $>$99.9\% of electron like 
signals. This proves that removing oxygen from the scintillator leads to 
improved particle identification and thus rejection of alphas. This is of 
particular importance in large scale liquid scintillator experiments such as 
SNO+. 

\section{Conclusions} 
This work has shown that the scintillator timing profile for linear 
alkylbenzene contains multiple decay components.  The relative 
amounts and values of these components are sufficiently different for electron 
and alpha events, that separation and thus discrimination between particle 
types is possible. By directly comparing measurements of oxygenated and 
deoxygenated scintillator, it has been demonstrated that removal of oxygen 
from the scintillator leads to improved particle identification and separation 
of alpha and electron like events.  While deoxygenating scintillator is 
widely expected to improve pulse shape discrimination, it is not commonly
demonstrated in the literature.  This work is the first to conclusively show
this is the case for linear alkylbenzene, a scintillator which will be used 
in several large scale neutrino physics experiments, including SNO+.
\section{Acknowledgements} 
The authors would like acknowledge financial support from the Natural Sciences 
and Engineering Research Council of Canada (NSERC).

\end{document}